# The ecology of social interactions in online and offline environments

Angelo Antoci [1]

Alexia Delfino [2]

Fabio Paglieri [3]

Fabio Sabatini [4]

**Abstract**

The rise in online social networking has brought about a revolution in social relations. However, its effects on offline interactions and its implications for collective well-being are still not clear and are under-investigated. We study the ecology of online and offline interaction in an evolutionary game framework where individuals can adopt different strategies of socialization. Our main result is that the spreading of self-protective behaviors to cope with hostile social environments can lead the economy to non-socially optimal stationary states.

**PsycINFO Codes**: 2240, 2750.

[1] Department of Economics and Business, University of Sassari, Italy. Email: antoci@uniss.it. The research of Angelo Antoci has been financed by Regione Autonoma della Sardegna (L. R. n. 7, 2007; research project *Capitale sociale e divari economici regionali*).

[2] Department of Economics, London School of Economics and Political Science, UK. Email: A.Delfino2@lse.ac.uk.

[3] National Research Council, Institute of Cognitive Sciences and Technologies, Italy. Email: fabio.paglieri@istc.cnr.it.

[4] Department of Economics and Law, Sapienza University of Rome, Italy Email: fabio.sabatini@uniroma1.it.

# 1. Introduction

The advent of social networking sites (SNS)[5] has caused a striking transformation in the way people interact with each other and with the surrounding social, political, and institutional environment. Online networks such as Facebook allow users to preserve and develop their social relations despite time and distance constraints (Ellison et al., 2007). SNS also provide users with strategies to cope with conditions of social decay: when the offline environment is lacking in social participation opportunities, individuals can turn to the online world or establish new contacts and create new chances of offline interaction (Steinfield et al., 2008). Situations of decay, however, can affect even online networks. Descriptive statistics from the Pew Research Center (PRC) illustrate that "online incivility", an important yet still under-investigated phenomenon - including aggressive and disrespectful behaviors, vile comments, harassment, and hate speech - is spreading across SNS and makes online social environments potentially hostile for users (Rainie et al., 2012; Duggan, 2014). Recent studies suggest that exposure to online incivility may be detrimental to SNS users' trust and well-being (Sarracino et al., 2015a; Sabatini et al., 2015).

Despite the extent of the transformations brought about by online networking, existing research on the relationships between online and offline interactions is limited. We still lack models to analyse the evolution of the different strategies of social engagement in potentially hostile environments and how it impacts collective well-being. This issue has important societal and economic implications, as social interaction influences the formation of opinions, political participation, collective action, and consumption patterns.

We study the ecology of online and offline social interaction in an evolutionary game framework where individuals can choose whether to be polite or not when meeting others in online environments. Everyone also has the option to isolate themselves from any social interaction, as a self-protective behavior to cope with hostile social environments. An offline environment can be hostile in relation to the decline in the opportunities for social engagement, due for example to increasing busyness and to the lack of meeting places such as parks, theatres, and associations. People may react by reducing face-to-face interaction, in order to develop part of their social life online, or even to seek refuge in social isolation, as suggested in Putnam's (2000) study on the decline of the American community. An online social environment, on the other hand, can become hostile with the spread of online incivility, which makes Internet-mediated interaction less rewarding. Users may want to defend themselves from incivility in social networks by adopting an uncivil behavior in their turn, or by abandoning those networks. In this latter case, if face-to-face interactions are perceived as not rewarding enough, self-protective behaviors may lead to social isolation.

---

[5] We use the terms social networking sites (SNS), online social networks and online networks as synonyms for the sake of brevity.

We define four types of agents pursuing different social interaction strategies: 1) people who develop social relations by exclusive means of face-to-face encounters. The distinctive trait of this type of agents is that they do not use SNS; 2) people who, in addition to face-to-face encounters, do use SNS and behave politely in online interactions; 3) SNS users who behave in an uncivil way in online interactions; 4) individuals who choose to withdraw from social relations. We define the equilibrium in which all individuals choose social isolation as a "social poverty trap" (Antoci et al., 2007; 2012a; 2015).

We model a homogeneous population, where every individual has the same preferences and access to technologies, but they differ with respect to the adopted strategy. However, the fourth strategy can also be interpreted as a form of self-protective behavior, which emerges when the combined hostility of physical and virtual social environmentsprompts a drastic form of adaptation consisting in the withdrawal from any form of social interaction (offline or online) and the choice of social isolation.

The analysis of dynamics shows that the spread of self-protective behaviors, as triggered by incivility and/or social poverty, entails undesirable results to the extent to which it leads the economy to non-socially optimal stationary states that are Pareto dominated by others. For individuals, self-protective behaviors are rational in that they temporarily provide higher payoffs. However, their spread causes a generalized decrease in the payoffs associated with each social participation strategy, which, in the long run, leads the economy to a non-optimal stationary state. The social poverty trap is always a locally attractive Nash equilibrium. When the other stationary states are attractive, they always give higher payoffs than the social poverty trap.

Our contribution bridges two literatures. The first literature is that of economists and political scientists who empirically analyzed how Internet use may impact on aspects of social capital such as face-to-face interactions and well-being (e.g. Falck et al., 2012; Campante et al., 2013; Helliwell and Huang, 2013; Bauernschuster et al., 2014). We contribute to this body of research by providing the first study of the possible evolution of various modes of offline and online interactions. In addition, we introduce the problem of online incivility and develop the first theoretical analysis of civility and incivility in online interactions.

Our focus on social poverty traps is also related to previous economic and sociological studies that analyzed how economic growth and technological progress may cause a decline in face-to-face social interactions (Putnam, 2000; Antoci et al., 2007; 2012; 2013; Bartolini and Bonatti, 2008; Bartolini and Sarracino, 2015), and to the literature concerning the 'decline of community life thesis' (Paxton, 1999, p. 88; Sarracino, 2010).

The second body of literature is that of psychologists and computer scientists who have analyzed the impact of SNS use on social capital and well-being (e.g. Ellison et al., 2007; Steinfield et al., 2008; Kross et al., 2013).

The paper begins by providing the motivation of the study and briefly reviewing the existing literature in sections 2, 3 and 4. We then illustrate the setup of the model. In section 6 we analyze the dynamics of the different forms of participation. Section 7 is devoted to a well-being analysis. In Section 8 we conclude by discussing some possible interpretations and policy implications of the results.

## 2. The decline in social engagement

In his best seller *Bowling Alone*, Robert Putnam (2000) documented that a decline in measures of social capital – such as participation in formal organisations, informal social connectedness, and interpersonal trust – began in the United States in the 1960s and 1970s, with a sharp acceleration in the 1980s and 1990s.

Putnam's 'decline of community life thesis' (Paxton, 1999, p. 88) prompted a number of subsequent empirical tests. Costa and Kahn (2003) used a number of different sources to assess the development of social capital in the United States since 1952. The authors found a decline in indicators of volunteering, membership of organisations and entertainment with friends and relatives. Based on GSS data, Bartolini et al. (2013) found a declining trend in indicators of social connectedness and confidence in institutions in the United States between 1975 and 2002.

Apart from the United States, there seems to be a common pattern of declining trust, social engagement and organisational activity across industrialised democracies starting from the 1980s, with the exception of Scandinavian countries (Leigh, 2003; Listhaug and Grønflaten, 2007). Declining trends of indicators of social interaction have been documented for England and Wales over the period 1972–99 (Li et al., 2003), Great Britain over 1959–90 (Hall, 2010) and 1980–2000 (Sarracino, 2010), China (Bartolini and Sarracino, 2015) and Australia over 1960–90 (Cox, 2002).

Putnam (2000) discussed three main explanations for the decline in American social capital: 1) the reduction in the time available for social interaction – related to the need to work more, to the rise in labour flexibility and to the increase in commuting time in urban areas; 2) the rise in mobility of workers and students; and 3) technology and mass media.

In the last decade, Putnam's arguments have found support in a number of studies investigating the effect exerted on various dimensions of social connectedness by the rise in working time (Bartolini and Bilancini, 2011), labour mobility (Routledge and von Ambsberg, 2003), urban sprawl and commuting (Besser et al., 2008; Wellman et al., 2001), and the impoverishment of the social environment, which can prompt individuals to pursue isolation (Bartolini and Bonatti, 2003; Antoci et al., 2007).

Antoci et al. (2012a; 2013) modelled the decline in social engagement as the result of a self-protective reaction to the reduction in the time available for social activities, the decline in social participation opportunities and the rise of materialistic values. According to the authors, the need to "defend" oneself from an unfriendly environment where social engagement becomes increasingly less rewarding prompts the substitution of relational goods with private goods in individual preferences, thereby favouring social isolation. Social isolation can be interpreted as a particular form of self-defense through which individuals make their utility independent from the actions of others. For example, individuals choosing social isolation tend to watch a movie alone through a home theatre system instead of going to the cinema with friends. They may even prefer to renounce their leisure activities to devote all of the available time to work. In this way, their payoffs do not vary with the closing of theatres or with the decline in the number (or even the unavailability) of friends with whom to share a night at the cinema. This shift in preferences is not driven by mutating tastes. Rather, as explained by Hirsch (1976), it must be interpreted as a self-protective reaction to the deterioration of the social environment. Hirsch (1976) was the first to introduce the concept of defensive consumption choices in his seminal work on the social limits to growth. This kind of consumption occurs in response to a change in the physical or social environment: "If the environment deteriorates, for example, through dirtier air or more crowded roads, then a shift in resources to counter these "bads" does not represent a change in consumer tastes but a response, on the basis of existing tastes, to a reduction in net welfare" (Hirsch, 1976, p. 63).

## 3. The rise in SNS-mediated interaction

In *Bowling Alone*, Putnam (2000) argued that progress in information technology could further exacerbate the decline in community life. At the time, Putnam referred to the negative role of television and other forms of technology-based entertainment such as video players and videogames. Early Internet studies reprised Putnam's arguments suggesting that the Internet might displace even more social activities than television (DiMaggio et al., 2001). The displacement hypothesis was supported by the first empirical tests of the relationship between Internet use and face-to-face interactions (e.g. Kraut et al., 1998; Nie et al., 2002).

These explorations, however, were carried out before the rise of SNS, when using the Internet was predominantly a solitary activity with limited relational implications. Today, Internet use is closely related to engagement in online social networks.

According to the Pew Research Center (PRC) Internet & American Life Project Survey, as of September 2014, 71 per cent of online adults were active on Facebook, 23 per cent used Twitter, 28 per cent used Pinterest and 26 per cent used Instagram (Duggan et al., 2015).

These figures mark a dramatic increase from 2009, when the PRC first began collecting data on Internet use. At that time, 46 per cent of online adults had ever used a SNS (Duggan and Brenner, 2013). Despite the extent of this transformation, the economic research on online networks is limited. In the fields of social psychology and communication science, several authors have tackled the potential role of SNS in face-to-face interaction. Ellison et al. (2007) studied how Facebook affected social capital and well-being in a sample of undergraduate students in an American college. They found a strong association between the use of Facebook and aspects of social capital entailing repeated and trust-based interactions with family and friends. In a longitudinal follow-up of the study, Steinfield et al. (2009) found that Facebook use in year one strongly predicted indicators of bridging social capital outcomes in year two. Based on a survey administered to college students recruited from two Texas universities, Valenzuela et al. (2013) found positive relationships between the intensity of Facebook use and students' life satisfaction, social trust, civic engagement, and political participation.

In economics, a few studies empirically assessed the role of broadband on aspects of social capital and political participation but, due to a lack of data, they could not tackle the possible role of online social networks. Based on German Socio-Economic Panel data, Bauernschuster et al. (2014) found that having broadband Internet access at home has positive effects on individuals' social interactions, manifesting in a higher frequency of visiting theatres, opera and exhibitions, and in a higher frequency of visiting friends. Using data on Italian municipalities, Campante et al. (2013) found that the diffusion of broadband led, initially, to a significant decline in electoral turnout in national parliamentary elections. This was reversed in the 2013 elections when the first round took place after the explosive rise of SNS. Falck et al. (2012) found that the progressive increase in DSL availability significantly decreased voter turnout in German municipalities.

Based on cross-sectional Italian data, Sabatini and Sarracino (2014a) suggested that online social networks may be used to preserve and consolidate existing relationships and that the use of SNS may play a role in preventing social isolation. The authors argued that the positive relationships between broadband availability and aspects of social capital identified in previous economic studies might, in fact, be due to the use of online social networks.

Antoci et al. (2012a; 2012b; 2015) theoretically analysed the evolution of social participation and the accumulation of social capital in relation to technological progress and online networking. Their results suggest that, under certain conditions, the stock of information and social ties accumulated within online networks can create an infrastructure that helps individuals to develop their social participation despite space and time constraints.

Overall, the evidence suggests that face-to-face and SNS-mediated interaction may be complementary, rather than one substituting the other. On the other hand, there is evidence that, despite the steep rise in the use of SNS, a decreasing yet still remarkable share of online adults chooses not to use them (see for example Zickuhr, 2013, for the U.S. and Sabatini and Sarracino, 2014a, for Italy). These facts point to the need for modeling a social interaction strategy exclusively based on face-to-face encounters and one entailing both face-to-face and SNS-mediated interactions.

## 4. The problem of online incivility

The rise of SNS-mediated interaction has been accompanied by the emergence of new, unexpected, downsides. Anecdotal and descriptive evidence shows that interaction in online social networks has increasingly been plagued by online incivility. The PRC reports that notable proportions of SNS users do witness bad behavior on those sites. 49% of SNS-using adults said they have seen mean or cruel behavior displayed by others at least occasionally, and 26% said they have had bad experiences that have caused face-to-face confrontations, problems with their family, or troubles at work (Rainie et al., 2012). According to a recent survey on online harassment, 73% of adult Internet users have seen someone harassed in some way in SNS, and 40% have personally experienced it. 60% of Internet users said they had witnessed someone being called offensive names in SNS, 53% had seen efforts to purposefully embarrass someone, 25% had seen someone being physically threatened, and 24% witnessed someone being harassed for a sustained period of time. "Fully 92% of Internet users agreed that the online environment allows people to be more critical of one another, compared with their offline experiences. Some 63% thought online environments allow for more anonymity than in their offline lives" (Duggan, 2014: p. 1). Women aged 18-24 are more likely than others to experience some of the more severe forms of harassment, including being called offensive names, stalked, and sexually harassed.

The roots of incivility in SNS-mediated social interactions have been addressed in a few psychological studies, which suggest that, when it comes to the presentation of opposing views and opinions, there is a fundamental difference between face-to-face and Internet-mediated interactions.

In contrast to online conversations, face-to-face interactions entail the use of expressions, smiles, eye contact, tone of voice, gesturing, and other nonverbal behavior that makes it easier to correctly perceive the interlocutors' feelings and intentions. Online conversations, on the other hand, are more vulnerable to incomprehension and misunderstandings. In SNS-mediated interactions, interlocutors are basically 'invisible' and their feelings and sensitivities can hardly be perceived. As stated by Kiesler et al. (1984) in an early study on computer-mediated communication, "Communicators must imagine their audience, for at a terminal it almost seems as though the computer itself is the audience. Messages are depersonalized, inviting stronger or more uninhibited text and more assertiveness in return". Kiesler et al. (1984) observed that computer-mediated communication entails anonymity, reduced self-regulation, and reduced self-awareness." The overall weakening of self- or normative regulation might be similar to what happens when people become less self-aware and submerged in a group, that is, *deindividuated*" (p. 1126). Deindividuation has in turn been found to be conducive to disinhibition and lack of restraint (Diener, 1979).

As a result, while in physical interactions people usually think twice before behaving offensively with a person who expresses an opposing view, SNS users are likely to care less about the risk of offending others in online conversations. In a pioneering experiment comparing face-to-face and online conversations, Siegel et al. (1983) found that people in computer-mediated groups were more aggressive than they were in face-to-face groups. In general, they were more responsive to immediate textual cues, more impulsive and assertive, and less bound by precedents set by societal norms of how to behave in groups. Based on survey data collected in a big U.S. company, Sproull and Kiesler (1986) found that computer-mediated communication has substantial deregulating effects and encourages disinhibition in respect to non-mediated interactions.

Further studies suggested that a more impulsive and assertive behavior that does not consider the recipients' feelings is far more common in Internet-mediated discussions as compared to face-to-face encounters (Lea et al., 1992; Orengo et al., 2010). This phenomenon has been conceptualized as "flaming" (Siegel et al., 1986). It refers to the expression of strong and uninhibited opinions, consisting of extreme emotional behavior expressed through uninhibited speech (insulting, offending, hostile comments, etc.).

The experimental studies mentioned above were conducted in very limited networks that were created ad hoc by researchers. It is reasonable to argue that in large networks such as Facebook and Twitter deindividuation and, therefore, disinhibition are likely to be exacerbated. Recent studies on Facebook have shown that controversial content was more frequent than any prosocial content categories, suggesting that there is an overrepresentation of negative content on the platform (Shelton and Skalski, 2014). A further distinctive element of interaction in big online networks is that possible reactions to provocative behaviours can be easily neutralized, for example by simply switching off the device, or even by 'blocking' the interlocutor through the network's privacy settings. These 'exit options' probably further weaken inhibitions and self-regulation. By contrast, one cannot easily withdraw from an unpleasant face-to-face discussion.

The problem of incivility is important because the infringement of social norms for the polite expression of opposing views can provoke emotional and behavioral responses with relevant economic and political consequences.

Mutz and Reeves (2005) experimentally analyzed the impact of incivility in mediated communication on trust. The authors noted a fundamental difference between face-to-face and television-mediated discussions about political views. Television-mediated presentations of opposing opinions often violate face-to-face social norms and easily deviate from civility. Mutz and Reeves (2005) collected experimental evidence that witnessing televised incivility causes a loss of trust in others. The authors claimed that, when social norms of politeness are violated in televised debates, watchers might feel hurt as if they personally experienced the offences they saw on TV. Sabatini and Sarracino (2015) argued that when incivility takes place in SNS-mediated interactions, users' feelings might be affected as if the offences where perpetrated in real life. In respect to what may happen with televised incivility, witnessing online incivility entails a more intense emotional involvement not only because one can be personally targeted with offensive behaviors but also because when others are being offended in online environments there is a concrete possibility to intervene in their defense. Based on Italian survey data, Sabatini and Sarracino (2014b; 2015) found that SNS users have significantly lower levels of trust in strangers, in neighbors, and in institutions than non-users and that such a decline in trust may be detrimental for users' well-being. The use of SNS could cause a decline in trust through different mechanisms, some of which have already been mentioned: for instance, increased awareness of diversity, experience of new social norms and more frequent exposure to incivility as compared to face-to-face interactions.[6]

Overall, the evidence regarding online incivility suggests that SNS can easily become a hostile environment for users and prompts the need to analyze two different strategies of social interaction via SNS, based on the propensity for acting civilly or not.

## 5. The model

Let us assume that, in each instant of time *t*, individuals play a one-shot population game (i.e. all individuals play the game simultaneously); they have to choose (ex ante) one of the following strategies:

***Strategy O***: social relations are developed by exclusive means of face-to-face encounters. We call this strategy *O* (for Offline). The distinctive trait of agents playing *O* is that they do not use online social networks.

***Strategy H***: social relations are developed both by means of face-to-face interactions and via social networking sites. SNS users who choose *H* (for Hate) behave online in an uncivil way. These agents indulge in offensive and disrespectful behaviors, vile comments, online harassment, or hate speech.

***Strategy P***: agents who follow this strategy develop their social relations both by means of face-to-face interactions and via SNS. In contrast to *H* players, however, *P* players behave politely in online interactions. We call this strategy *P* (for Polite).

---

[6]Accordingly, in the model we make the simplifying assumption that people only behave politely when interacting *offline*. On the one hand, the very nature of face-to-face interactions increases the costs of incivility, reducing the incentives to behave impolitely offline as compared to the Internet. On the other, given the proportion of people adopting each strategy, in each period people interacting online meet more people than those who are only offline.

The *H* and *P* modes of participation entail different degrees of Internet-mediated interaction according to individuals' preferences: in general, we think of SNS users as individuals who develop social ties online at their convenience—for example, by staying in touch with friends and acquaintances, or interacting with unknown others online—and meet their contacts in person whenever they want or have time.

**Strategy *N*:** finally, we assume that individuals can choose to withdraw from social relations by reducing them to the minimum, in order to devote all their time to private activities such as work or material consumption. We label this strategy as *N* (for No social participation) and we call the equilibrium in which all individuals withdraw from social participation a "social poverty trap" (Antoci et al., 2007; 2013; 2015). The withdrawal from social interactions modeled with the *N* strategy may be viewed as a drastic form of adaptation to conditions of social decay that make *N* players' payoff constant and completely independent from the behavior of others.

Let us indicate by $x_1(t), x_2(t), x_3(t), x_4(t)$ the shares of individuals adopting strategies *O, H, P,* and *N*, respectively, at time *t*. It holds $x_i \geq 0$, all *i*, and $\sum_i x_i = 1$, therefore the vector $(x_1, x_2, x_3, x_4)$ belongs to the three-dimensional simplex *S* represented in Figure 1.

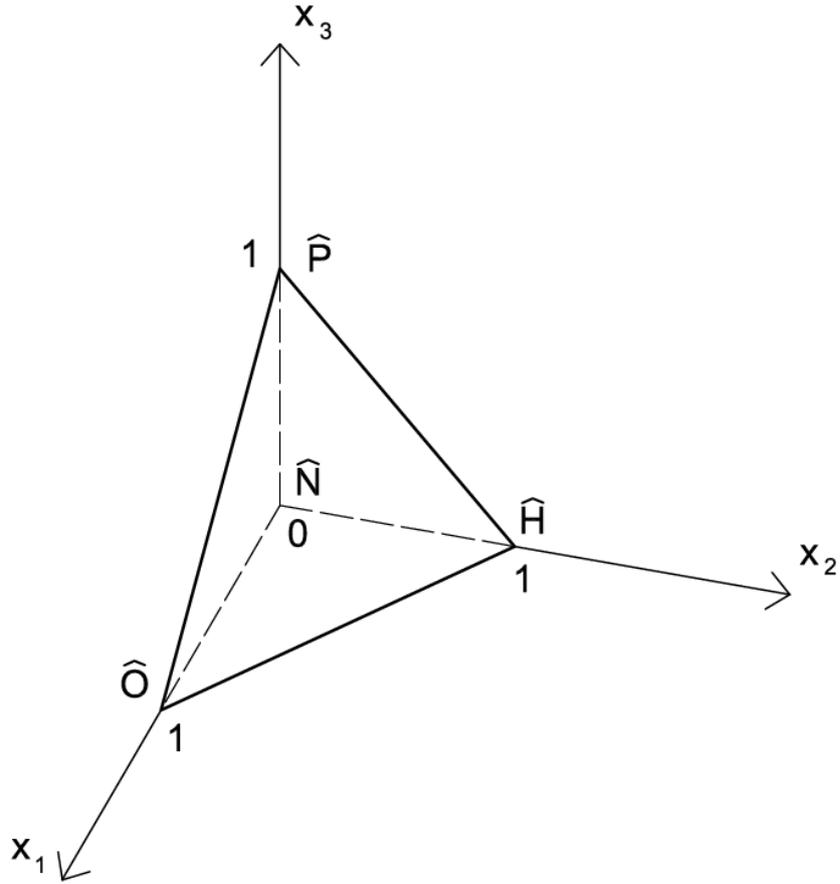

**Figure 1**: The three-dimensional simplex $S$ in the space $(x_1, x_2, x_3)$. The value of $x_4$ is given by the equation $x_4 = 1 - x_1 - x_2 - x_3$ and the origin $(0,0,0)$ of the space $(x_1, x_2, x_3)$ corresponds to the vertex $\hat{N}$ of $S$ where $x_4 = 1$. In the other vertices $\hat{O}$, $\hat{H}$, $\hat{P}$ all individuals play, respectively, strategies $O$, $H$, $P$.

For simplicity, we assume that the payoff functions of the strategies $O$, $H$ and $P$ are linear in the variables $x_1$, $x_2$, and $x_3$:

$$\Pi_O(x_1) = \alpha \cdot x_1$$

$$\Pi_H(x_2, x_3) = \beta \cdot x_2 + \gamma \cdot x_3$$

$$\Pi_P(x_2, x_3) = -\delta \cdot x_2 + \varepsilon \cdot x_3$$

While the payoff of the *N* strategy is constant and, therefore, does not depend on the distribution $(x_1, x_2, x_3, x_4)$ of strategies:

$$\Pi_N = \eta$$

We assume $\alpha, \delta, \varepsilon, \eta > 0$. The strict positivity of $\eta$ characterizes *N* as a self-protective strategy: in a context where no one engages in social interactions, *N* becomes the best performing strategy. We also assume that the payoff from virtuous social interactions (i.e. adopting strategies *O* or *P*) is increasing in the proportion of people interacting in such a way ($\alpha$ and $\varepsilon$ positive). We assume the impact of the diffusion of the "hate" strategy on a polite's payoff is always negative ($\delta$ positive).[7]

We instead allow the parameters $\beta$ and $\gamma$ to be either positive or negative. It is not clear, in fact, whether haters get more satisfaction when dealing with polite SNS users or by confronting with others of the same type. A *H* player, for example, may find the interaction with a polite player who defuses provocations with kindness less rewarding; accordingly, we allow *H* players to even get disutility from the interaction with a polite person. Or, by contrast, she may find it harder, and less rewarding, to confront another hater.

For the sake of simplicity we also assume that individuals who only interact offline (i.e. those who choose the *O* strategy) never meet those who choose the mixed strategies entailing both face-to-face and online interaction (the *P* and the *H* strategy).[8]

The strategic context of the game can be better illustrated by the following payoff matrix:

|   | *All play O* | *All play H* | *All play P* | *All play N* |
|---|---|---|---|---|
| O | $\alpha$ | 0 | 0 | 0 |
| H | 0 | $\beta$ | $\gamma$ | 0 |
| P | 0 | $-\delta$ | $\varepsilon$ | 0 |
| N | $\eta$ | $\eta$ | $\eta$ | $\eta$ |

---

[7] Alternatively, we could assume $\delta$ to be negative if an increase in the proportion of haters increases the payoff for polite people, given the share of the population using the *P* strategy. While this can be realistic when $\gamma$ is also negative (so that haters' payoff is decreasing in the proportion of polite people), it is however harder to justify when $\gamma$ is positive.

[8] Participation in SNS can be considered one of the individual characteristics people share within their social network, either through assortative matching in the network itself or increasing similarity over time. It also makes sense intuitively: people would probably prefer to have "online experiences" to share if they are surrounded by friends who all have Facebook accounts and frequently talk about what happens there.

representing the payoffs of an individual playing the *O, H, P,* and *N* strategies "against" homogeneous populations where all individuals play only one strategy (either *O, H, P,* or *N*). Notice that:

**1)** The population state $\hat{N} = (x_1, x_2, x_3, x_4) = (0,0,0,1)$ -where all individuals play the *N* strategy - is always a (strict) Nash equilibrium.

**2)** The population state $\hat{O} = (x_1, x_2, x_3, x_4) = (1,0,0,0)$ -where all individuals play the *O* strategy- is a Nash equilibrium if and only if $\alpha > \eta$.

**3)** The population state $\hat{H} = (x_1, x_2, x_3, x_4) = (0,1,0,0)$ -where all individuals play the *H* strategy- is a Nash equilibrium if and only if $\beta > \eta$.

**4)** The population state $\hat{P} = (x_1, x_2, x_3, x_4) = (0,0,1,0)$ -where all individuals play the *P* strategy - is a Nash equilibrium if and only if $\varepsilon > \gamma, \eta$.

**5)** The pure population states $\hat{N}$, $\hat{O}$, $\hat{H}$, and $\hat{P}$ can simultaneously be Nash equilibria.

**6)** The payoff of each individual in the state $\hat{N}$ (given by $\eta$) is lower than the payoff of each individual in the states $\hat{O}$, $\hat{H}$, and $\hat{P}$ (given, respectively, by $\alpha$, $\beta$, and $\varepsilon$) when such states are Nash equilibria.

**7)** The *N* strategy is never dominated by the other strategies. The *O* strategy is dominated by the *N* strategy if $\alpha \leq \eta$, while no dominance relationship can occur between the *O* and the *H* strategy and between the *O* and the *P* strategy. The *H* strategy is dominated by *N* if $\eta \geq \max(\beta, \gamma)$, while it is dominated by *P* if $\beta \leq -\delta$ and $\gamma \leq \varepsilon$. Finally, the *P* strategy is dominated by *N* if $\varepsilon \leq \eta$, while it is dominated by *H* if $\beta \geq -\delta$ and $\gamma \geq \varepsilon$.

According to a well-known result in evolutionary game theory (see, e.g., Weibull, 1995), if the pure population states $\hat{N}$, $\hat{O}$, $\hat{H}$, and $\hat{P}$ are Nash equilibria, then they also are (locally) attractive stationary states under every payoff-monotonic adoption dynamics of strategies. Consequently, in the contexts in which $\hat{N}$ is not the unique existing Nash equilibrium, the adoption dynamics are path dependent in that different stationary states may be reached starting from different initial distributions $(x_1(0), x_2(0), x_3(0), x_4(0))$ of strategies.

The stationary state $\hat{N}$ can be interpreted as a *social poverty trap*, in the sense of Antoci et al. (2007); that is, as an attractive stationary state where aggregate social participation and welfare (measured by payoffs) fall to the lowest possible level with respect to other stationary states.

To focus our analysis on more relevant cases only, we shall study adoption dynamics under the assumption that no strategy is dominated by others (see Point 7 above). Such assumption requires the following restrictions on parameters' values:

$$\alpha > \eta, \varepsilon > \eta, \max(\beta, \gamma) > \eta \tag{1}$$

$$\text{either } \beta > -\delta \text{ and } \gamma < \varepsilon \text{ or } \beta < -\delta \text{ and } \gamma > \varepsilon \tag{2}$$

Notice that under condition $\alpha > \eta$ (see (1)), the state $\hat{O} = (x_1, x_2, x_3, x_4) = (1,0,0,0)$ -where all individuals play the $O$ strategy- is always a Nash equilibrium and, therefore, a locally attractive stationary state. So, at least two locally attractive stationary states - $\hat{O}$ and the *social poverty trap* $\hat{N}$ - exist under every payoff-monotonic adoption dynamics. Furthermore:

a) In the context in which $\beta > -\delta$ and $\gamma < \varepsilon$ hold (see (2)), then also the state $\hat{P} = (x_1, x_2, x_3, x_4) = (0,0,1,0)$ -where all individuals play the $P$ strategy- is a Nash equilibrium and a locally attractive stationary state, while the state $\hat{H} = (x_1, x_2, x_3, x_4) = (0,1,0,0)$ -where all individuals play the $H$ strategy- may be a Nash equilibrium (this is the case only if $\beta > \eta$) or not.

b) In the context in which $\beta < -\delta$ and $\gamma > \varepsilon$ hold (see (2)), then the states $\hat{P}$ and $\hat{H}$ are never Nash equilibria (and, therefore, they are never attractive). As we will see, such context favours the coexistence of the $H$ and the $P$ strategy. Importantly, this context captures an interesting set of social scenarios: the first condition, $\beta < -\delta$, requires that a $H$ player is more negatively affected by interacting with another $H$ player than what would happen to a $P$ player, suggesting that (i) haters do not get along with each other, possibly because they get no satisfaction in the absence of a proper "victim" and/or are forced to take a taste of their own medicine, whereas (ii) polite people are only mildly annoyed by interacting with haters. On the other hand, the second condition, $\gamma > \varepsilon$, implies that interacting with a $P$ player is more satisfactory for a $H$ player than for another $P$ player. Taken together, these conditions thus describe a context in which haters have stronger "online passions" than polite participants, that is, partner selection is even more crucial for haters than for polite users – which seems a plausible characterization of many real-life interaction scenarios.

## 6. Evolutionary dynamics

Following Taylor and Yonker (1978), we assume that the diffusion of the four strategies is described by the replicator equations:

$$\begin{aligned}
\dot{x}_1 &= x_1[\Pi_O(x_1) - \overline{\Pi}(x_1, x_2, x_3, x_4)] \\
\dot{x}_2 &= x_2[\Pi_H(x_2, x_3) - \overline{\Pi}(x_1, x_2, x_3, x_4)] \\
\dot{x}_3 &= x_3[\Pi_P(x_2, x_3) - \overline{\Pi}(x_1, x_2, x_3, x_4)] \\
\dot{x}_4 &= x_4[\Pi_N - \overline{\Pi}(x_1, x_2, x_3, x_4)]
\end{aligned} \quad (3)$$

Where $\dot{x}_i = dx_i(t)/dt$ represents the time derivative of $x_i$, $i=1,...,4$, and:

$$\overline{\Pi}(x_1, x_2, x_3, x_4) = x_1 \cdot \Pi_O(x_1) + x_2 \cdot \Pi_H(x_2, x_3) + x_3 \cdot \Pi_P(x_2, x_3) + x_4 \cdot \Pi_N$$

is the population-wide average payoff.

According to replicator equations (3), individuals tend to imitate players who adopt the relatively more rewarding strategies. As a consequence, such strategies spread in the population at the expenses of the less rewarding ones.

Replicator dynamics (3) are defined in the three-dimensional simplex:

$$S = \left\{ (x_1, x_2, x_3, x_4) \in R^4 : x_i \geq 0 \text{ all } i, \sum_i x_i = 1 \right\}$$

represented in Figure 1 in the space $(x_1, x_2, x_3)$.[9] All the pure population states $\hat{N}$, $\hat{O}$, $\hat{H}$, and $\hat{P}$ are stationary states under dynamics (3). Furthermore, the edges of $S$ where one or more strategies are adopted by no one are invariant under dynamics (3); that is, every trajectory starting from a point belonging to one of the edges, remains in the edge for every time $t \in (-\infty, +\infty)$.

---

[9] The value of $x_4$ is given by the equation $x_4 = 1 - x_1 - x_2 - x_3$ and the origin $(0,0,0)$ of the space $(x_1, x_2, x_3)$ corresponds to the point of $S$ where $x_4 = 1$.

To analyze replicator dynamics (3), it is useful to take into account the well-known correspondence between replicator equations and Lotka-Volterra systems (Hofbauer, 1981). In particular, in this case, we have that the transformation:

$$T: (y, z, w) \to (x_1, x_2, x_3, x_4) =$$

$$= \left( \frac{1}{1+y+z+w}, \frac{y}{1+y+z+w}, \frac{z}{1+y+z+w}, \frac{w}{1+y+z+w} \right)$$

maps the trajectories under Lotka-Volterra equations:

$$\begin{aligned} \dot{y} &= y(-\alpha + \beta y + \gamma z) \\ \dot{z} &= z(-\alpha - \delta y + \varepsilon z) \\ \dot{w} &= w[-\alpha + \eta(1 + y + z + w)] \end{aligned} \quad (4)$$

onto those generated by replicator equations (3). The inverse transformation of $T$:

$$T^{-1}: (x_1, x_2, x_3, x_4) \to (y, z, w) = \left( \frac{x_2}{x_1}, \frac{x_3}{x_1}, \frac{x_4}{x_1} \right)$$

does the opposite.

Variables $y$, $z$, and $w$ respectively measure the ratios between the shares $x_2$, $x_3$, $x_4$ of individuals playing the $H$, $P$, and $N$ strategy and the share $x_1$ of individuals playing the $O$ strategy.

According to system (4), the paths followed by the ratios $y = x_2/x_1$ and $z = x_3/x_1$ do not depend on the ratio $w = x_4/x_1$. This implies that the process of diffusion of strategies $O$, $H$, and $P$ in the sub-population of individuals not adopting the $N$ strategy does not depend on the ratio $w = x_4/x_1$ and is fully described by the trajectories generated by the sub-system of (4):

$$\dot{y} = y(-\alpha + \beta y + \gamma z)$$
$$\dot{z} = z(-\alpha - \delta y + \varepsilon z)$$
(5)

Obviously, the study of system (5) gives information about the ratios between the variables $x_2$, $x_3$ and the variable $x_1$, but it does not allow us to get information about the ratio $w = x_4/x_1$ and, therefore, about the ratio between the size of the sub-population of individuals choosing to *not participate* (i.e. playing the *N* strategy) and the size of the sub-population of individuals choosing to *participate* (i.e. choosing one of the strategies *O*, *H*, *P*).

### 6.1. The dynamics of social participation strategies: O, H, and P

In this section, we analyze system (5). The coordinate change proposed by Hofbauer (1981):

$$T:(y,z) \to (x_1, x_2, x_3) = \left( \frac{1}{1+y+z}, \frac{y}{1+y+z}, \frac{z}{1+y+z} \right)$$

maps the trajectories generated by the Lotka-Volterra system (5) onto those generated by the replicator equations:

$$\dot{x}_1 = x_1 [\Pi_O(x_1) - \overline{\Pi}(x_1, x_2, x_3)]$$
$$\dot{x}_2 = x_2 [\Pi_H(x_2, x_3) - \overline{\Pi}(x_1, x_2, x_3)]$$
$$\dot{x}_3 = x_3 [\Pi_P(x_2, x_3) - \overline{\Pi}(x_1, x_2, x_3)]$$
(6)

obtained by posing $x_4 = 0$ in system (3). The inverse transformation of *T*:

$$T^{-1}:(x_1, x_2, x_3) \to (y, z) = \left( \frac{x_2}{x_1}, \frac{x_3}{x_1} \right)$$

does the opposite.

System (6) describes the dynamics of system (3) in the two-dimensional edge with $x_4 = 0$ (where nobody plays the *N* strategy) of the three-dimensional simplex *S* represented in Figure 1. However, since the trajectories of system (6) are associated to the trajectories of the two-dimensional Lotka-Volterra system (5) via the coordinate change *T*, some insights obtained from the analysis of (6) can be transferred to (3), as we will show below.

The analysis of system (6) builds on the classification results in Bomze (1983) and is reported in the mathematical appendix. It allows us to give a complete classification of all the possible dynamic regimes that may be observed under system (6). These regimes are illustrated in Figures 2a-2f. In these and in the following figures, a full dot • represents a locally attractive stationary state, an empty dot ○ represents a repulsive stationary state, while a saddle point is indicated by drawing its insets and outsets (stable and unstable manifolds, respectively). Only some representative trajectories are sketched. The dynamic regimes that may be observed in the edge $S_N$ are the following:

**1)** Case $\beta > 0$ (and therefore $\beta > -\delta$) and $\gamma < \varepsilon$. In this case, all the stationary states $\hat{O}$, $\hat{H}$, and $\hat{P}$ are locally attractive.[10] No other attractive stationary state exists. Figures 2a and 2b illustrate, respectively, the sub-cases in which condition:

$$\beta\varepsilon + \gamma\delta > 0 \qquad (7)$$

holds and the sub-case in which the opposite of condition (7) holds. They correspond, respectively, to phase portraits number 7 and 35 in Bomze's classification (from now on, we shall indicate the *phase portrait number # of Bomze's classification* with the symbol *PP#*).[11] To ease interpretation, condition (7) can be also expressed as $\gamma/\beta > -\varepsilon/\delta$. This inequality compares the ratio of marginal return when meeting a polite or a hater for Haters versus Polites (abusing terminology, as a sort of Marginal Rate of Substitution of Haters vs. Polites). If (7) holds, at the margin the rate at which haters are willing to forgo meeting one hater for meeting a Polite is greater than the rate for Polites.

**2)** Case $\beta < 0$ and $\gamma < \varepsilon$ (and therefore $\beta > -\delta$, by conditions (2)). In this case, only the stationary states $\hat{O}$ and $\hat{P}$ are locally attractive. Figures 2c and 2d illustrate, respectively, the sub-case in which condition (7) and the opposite of condition (7) hold (they correspond, respectively, to *PP*9 and *PP*37).

---

[10] A stationary state may be not attractive for all the trajectories belonging to the simplex *S*, although it is attractive for all the trajectories belonging to an edge of *S*.

[11] For simplicity, in this classification and in the subsequent ones, we omit consideration of non robust dynamic regimes, that is, the regimes observed only if an equality condition on parameters' values holds.

**3)** Case $\beta < 0$ and $\gamma > \varepsilon$ (and therefore $\beta < -\delta$, by conditions (2)).[12] In this case, the vertex $\hat{O}$ is always locally attractive.

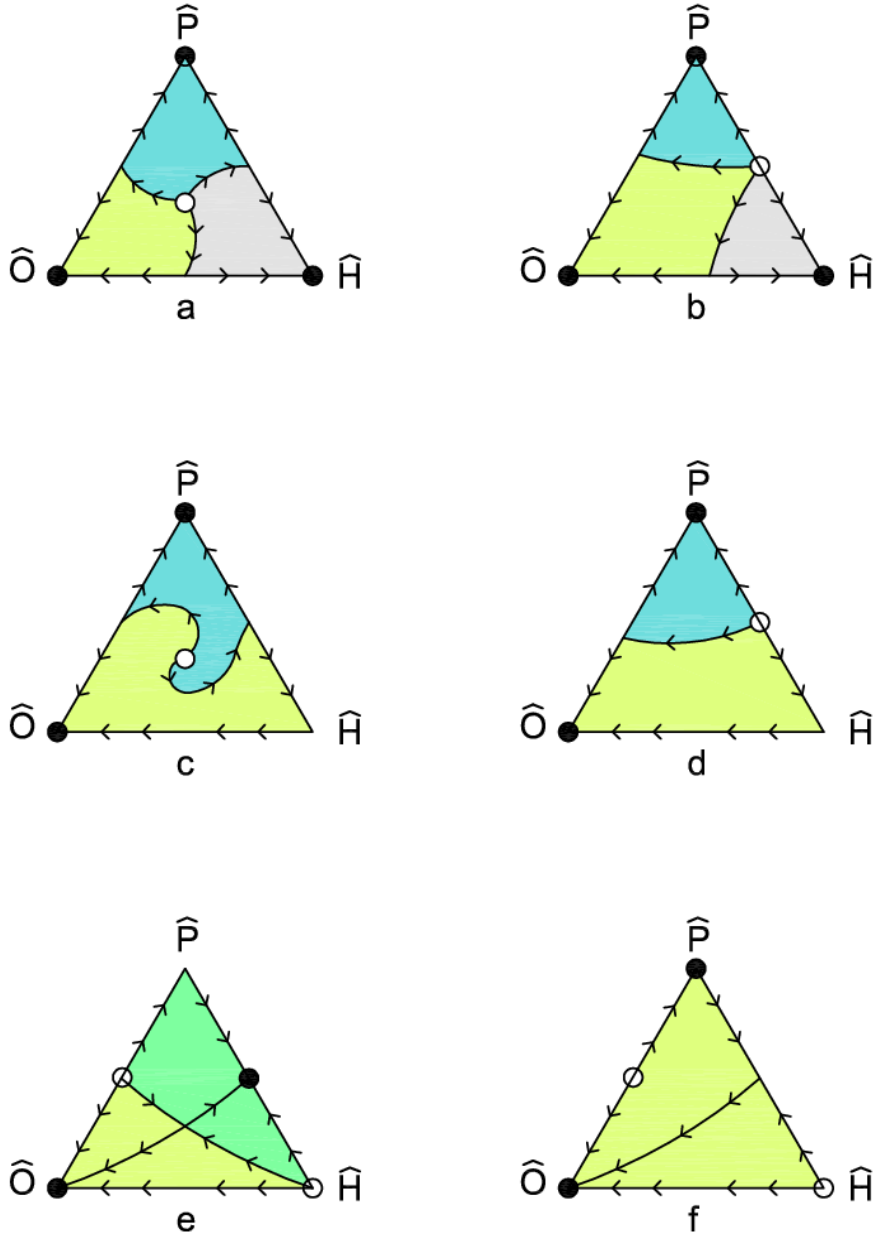

**Figure 2**: The dynamic regimes that may be observed in the edge $S_N$ of the simplex $S$, where the $N$ strategy is not played. In the stationary states $\hat{O}$, $\hat{H}$, $\hat{P}$ all individuals play, respectively, strategies $O$, $H$, $P$. In these and in the following figures, a full dot • represents a locally attractive stationary state, an empty dot ○ represents a repulsive stationary state, while a saddle point is indicated by drawing its insets and outsets. Only some representative trajectories are sketched.

---

[12] The case $\beta > 0$ and $\gamma > \varepsilon$ is excluded by conditions (2).

Furthermore, if the opposite of condition (7) holds, then there also exists another locally attractive stationary state lying in the edge of the simplex where only the *H* and the *P* strategies are played (see Figure 2e, corresponding to *PP*11), while if condition (7) holds, then no other attractive stationary state exists (see Figure 2f, which corresponds to *PP*36).

Notice that individuals' payoffs in the pure population stationary states $\hat{O}$, $\hat{H}$, and $\hat{P}$ are given respectively by $\Pi_O(1) = \alpha$, $\Pi_H(1,0) = \beta$, and $\Pi_P(0,1) = \varepsilon$; therefore, the stability conditions concerning such states (remember that $\hat{O}$ is always attractive, $\hat{H}$ is attractive if $\beta > 0$, $\hat{P}$ is attractive if $\gamma < \varepsilon$) do not allow us to order $\hat{O}$, $\hat{H}$, and $\hat{P}$ in terms of welfare. In addition, notice that in the stationary state where only the strategies *H* and *P* are played, we have $x_1 = x_4 = 0$, $x_3 = 1 - x_2$, $x_2 = x_2^* := (\varepsilon - \gamma)/(\beta + \delta + \varepsilon - \gamma)$, and individuals' payoff is given by:

$$\Pi_H(x_2^*, 1 - x_2^*) = \Pi_P(x_2^*, 1 - x_2^*) = \frac{\beta\varepsilon + \gamma\delta}{\beta + \delta + \varepsilon - \gamma} \tag{8}$$

If the stationary state $(x_1, x_2, x_3, x_4) = (0, x_2^*, 1 - x_2^*, 0)$ is attractive, then:

$$\varepsilon = \Pi_P(0,1) > \Pi_H(x_2^*, 1 - x_2^*) = \Pi_P(x_2^*, 1 - x_2^*) > \Pi_H(1,0) = \beta$$

holds: individuals' welfare in $(0, x_2^*, 1 - x_2^*, 0)$ is lower than in the (non-attractive) stationary state $\hat{P}$ and higher that in the (non-attractive) stationary state $\hat{H}$. Finally, the stationary states $(x_1, x_2, x_3, x_4) = (0, x_2^*, 1 - x_2^*, 0)$ and $\hat{O}$ cannot be ordered in terms of welfare, when they are both attractive (see Figure 2e).

As said above, the dynamic regimes illustrated in Figures 2a-2f are those that can be observed in the edge with $x_1 + x_2 + x_3 = 1$ (in correspondence of which, therefore, $x_4 = 0$ holds) of the simplex *S*. However, they also illustrate all the possible evolution paths that can be followed in the interior of the simplex *S* by the shares $x_1$, $x_2$, and $x_3$ of the sub-population composed by the individuals adopting social participation strategies (*O*, *H*, *P*). Consequently, the following insights can be learned from the dynamic regimes illustrated in Figures 2a-2f:

a) Under dynamics (3), the shares $x_1$, $x_2$, and $x_3$ always tend to stationary values (and, therefore, the share $x_4$ does it too); this implies that all the trajectories of system (3) tend asymptotically to a stationary state.

b) Attractive stationary states cannot exist where all the social participation strategies O, H, and P are adopted. This implies that in attractive stationary states of system (3), at most three strategies can coexist (among the available strategies O, H, P, and N) and, therefore, every attractive stationary state belongs to one of the edges of the simplex S (where at least one strategy is not adopted). Even if some trajectories may tend to a (non-attractive) stationary state where four strategies coexist,[13] "almost all" the trajectories tend to a stationary state belonging to one of the edges of the simplex S.

Since, according to the considerations above, almost all the trajectories of system (3) tend asymptotically to a stationary state belonging to one edge of the simplex S, it is useful to analyze the dynamics (3) in the edges of the simplex S. We do this in the following section.

### *6.2. Dynamics in the edges of the simplex S where the N strategy is played*

In order to avoid a lengthy presentation of our mathematical results, we omit the computations for classifying the dynamic regimes that may be observed in the remaining edges $S_O$, $S_P$, and $S_H$ of the three-dimensional simplex S where, respectively, strategies O, P, and H are not adopted. The procedure allowing us to apply Bomze's classification to such cases is very similar to that developed in the mathematical appendix to analyze the dynamics in the edge $S_N$.

### *6.2.1. Dynamics in the edge $S_O$*

The dynamic regimes that may be observed in the edge $S_O$ (where the O strategy is not played) are the following (see Figures 3a-3f):

---

[13] Notice that system (4) admits (generically) at most one stationary state $(\bar{y}, \bar{x}, \bar{w})$ with $\bar{y}, \bar{x}, \bar{w} > 0$. At $(\bar{y}, \bar{x}, \bar{w})$, all the four available strategies coexist. The Jacobian matrix of system (4), evaluated at $(\bar{y}, \bar{x}, \bar{w})$ has one strictly positive eigenvalue equal to $\eta\bar{w}$, while the other two eigenvalues coincide with those of the Jacobian matrix of system (5) evaluated at $(\bar{y}, \bar{x},)$. Since in the dynamic regime showed in Figure 2e the Jacobian matrix of system (5) has two eigenvalues of opposite signs, in such a context the Jacobian matrix of system (4), evaluated at $(\bar{y}, \bar{x}, \bar{w})$ has one negative and two positive eighenvalues. Consequently, there exists a one-dimensional stable manifold of $(\bar{y}, \bar{x}, \bar{w})$ and, therefore, there exist trajectories tending asymptotically to the (unstable) stationary state $(\bar{y}, \bar{x}, \bar{w})$.

**1)** Case $\gamma < \varepsilon$ (and therefore, by assumption (2), $\beta > -\delta$). We have two sub-cases:

1.a) If $\beta > \eta$, then all the stationary states $\hat{H}$, $\hat{P}$, and $\hat{N}$ are locally attractive. No other attractive stationary state exists. Figures 3a and 3b illustrate the corresponding dynamics regimes (respectively, *PP*7 and *PP*35).

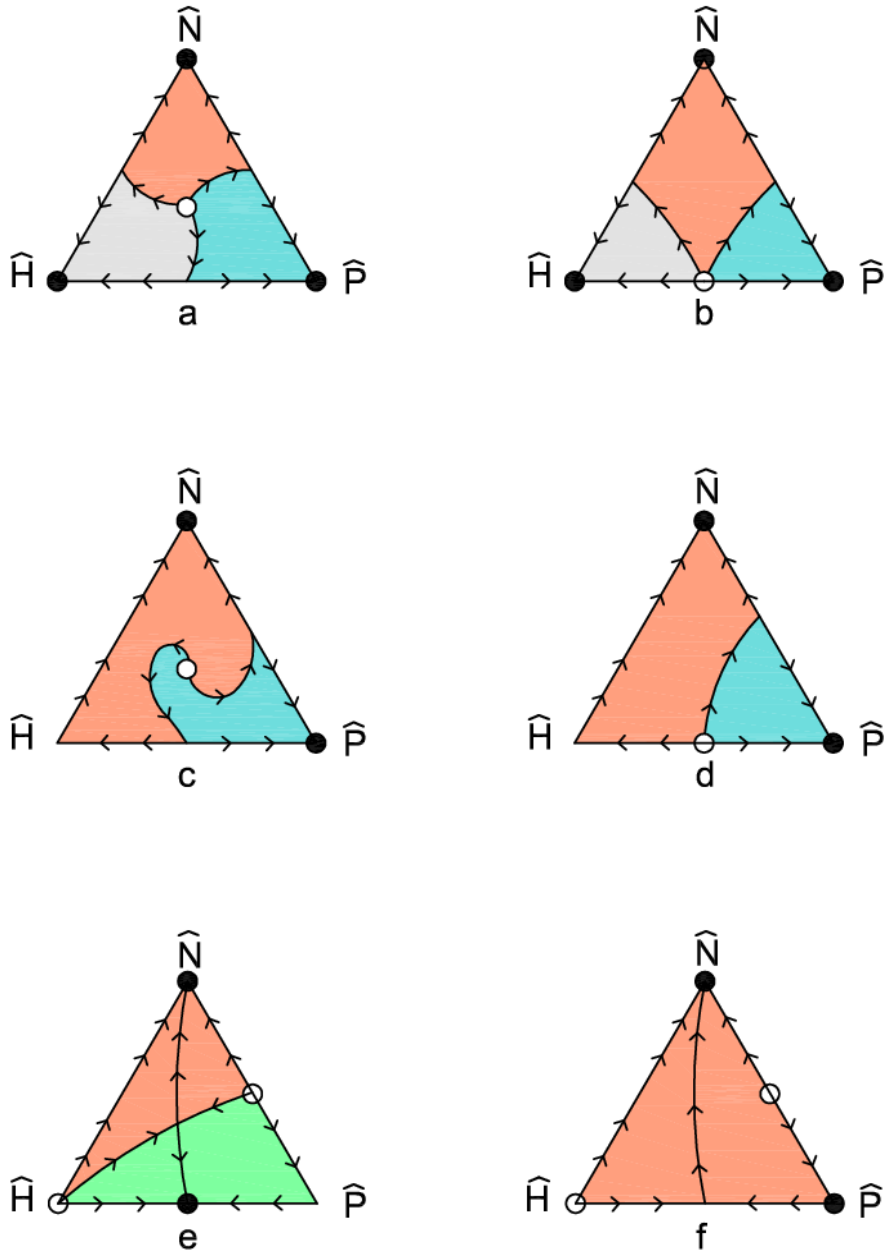

**Figure 3**: The dynamic regimes that may be observed in the edge $S_O$ of the simplex $S$, where the $O$ strategy is not played. In the stationary states $\hat{H}$, $\hat{P}$, $\hat{N}$ all individuals play, respectively, strategies *H*, *P*, *N*.

The regime in Figure 3a is observed if the condition $(\varepsilon - \gamma)(\eta - \beta) + (\beta + \delta)(\eta - \gamma) < 0$, that is:

$$\eta < \frac{\beta\varepsilon + \gamma\delta}{\varepsilon - \gamma + \beta + \delta} \qquad (9)$$

holds. The regime in Figure 3b is observed if the opposite of condition (9) holds.

1.b) If $\beta < \eta$, then the stationary states $\hat{P}$ and $\hat{N}$ are locally attractive, while $\hat{H}$ is a saddle point. Figure 3c (respectively, 3d) illustrates the dynamic regime occurring if condition (9) (respectively, the opposite of(9)) holds. Figures 3c and 3d correspond, respectively, to $PP9$ and $PP37$.

2) Case $\gamma > \varepsilon$ (and therefore, by assumption (2), $\beta < -\delta$). In this case, $\beta < 0$ holds and, therefore, $\beta < \eta$. According to these conditions on parameters, the stationary state $\hat{N}$ is locally attractive, $\hat{H}$ is repulsive and $\hat{P}$ is a saddle point. If condition (9) holds, then there exists another locally attractive stationary state $(0, x_2^*, 1 - x_2^*, 0)$ lying in the edge of the simplex where only strategies $H$ and $P$ are played (Figure 3f, which corresponds to $PP11$). If the opposite of condition (9) holds, then $\hat{N}$ is the unique attractive stationary state and the dynamic regime is that illustrated in Figure 3g (corresponding to $PP36$).

Notice that condition (9) holds if and only if (see (8)):

$$\Pi_H(x_2^*, 1 - x_2^*) = \Pi_P(x_2^*, 1 - x_2^*) > \Pi_N = \eta$$

This implies that when the stationary state $(x_1, x_2, x_3, x_4) = (0, x_2^*, 1 - x_2^*, 0)$ -where only strategies $H$ and $P$ are played- is attractive (see Figure 3e), then individuals' welfare in $(0, x_2^*, 1 - x_2^*, 0)$ is higher than in the social poverty trap $\hat{N}$.

### 6.2.2. Dynamics in the edge $S_H$

The dynamic regimes that may be observed in the edge $S_H$ (where the $H$ strategy is not played) are those illustrated in Figures 4a and 4b (corresponding, respectively, to $PP7$ and PP35). In both regimes, all the stationary states $\hat{O}$, $\hat{P}$, and $\hat{N}$ are locally attractive. No other attractive stationary state exists. The regime in Figure 4a is observed if the condition:

$$\eta < \frac{\alpha\varepsilon}{\alpha+\varepsilon} \qquad (10)$$

holds, while that in Figure 4b occurs if the opposite of condition (10) is satisfied.

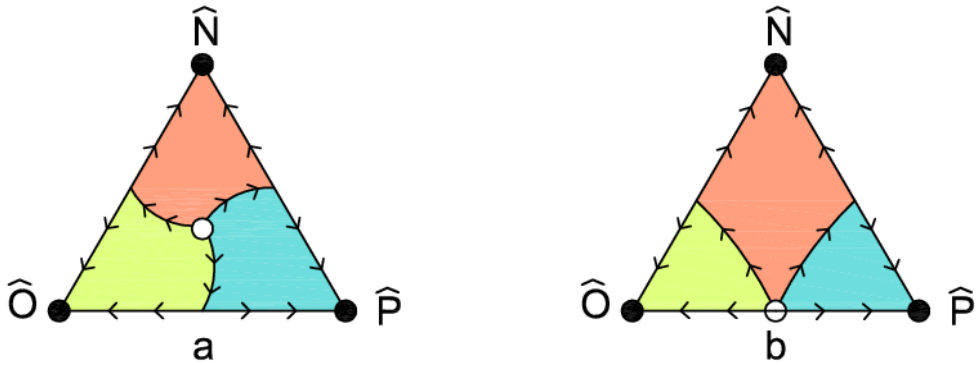

**Figure 4**: The dynamic regimes that may be observed in the edge $S_H$ of the simplex $S$, where the $H$ strategy is not played. In the stationary states $\hat{O}$, $\hat{P}$, $\hat{N}$ all individuals play, respectively, strategies $O$, $P$, $N$.

### 6.2.3. Dynamics in the edge $S_P$

The dynamic regimes that may be observed in the edge $S_P$ (where no one plays the $P$ strategy) are the following:

**1)** Case $\beta > \eta$. In this case, all the stationary states $\hat{O}$, $\hat{H}$, and $\hat{N}$ are locally attractive. No other attractive stationary state exists. Figures 5a and 5b illustrate, respectively, the sub-case in which condition:

$$\eta < \frac{\alpha\beta}{\alpha+\beta} \qquad (11)$$

holds and the sub-case in which the opposite of condition (11) holds. They correspond, respectively, to *PP*7 and *PP*35.

**2)** Case $\beta < \eta$. In this case, only the stationary states $\hat{O}$ and $\hat{N}$ are locally attractive. Figure 5c shows the corresponding dynamic regime (*PP*37).

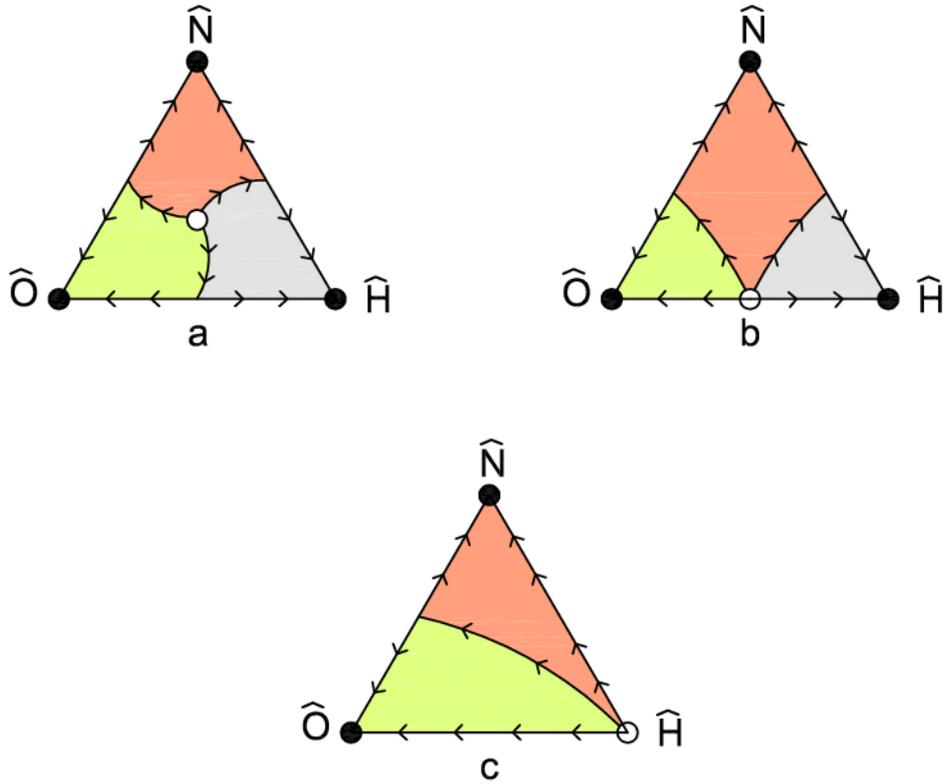

**Figure 5:** The dynamic regimes that may be observed in the edge $S_P$ of the simplex $S$, where the $P$ strategy is not played. In the stationary states $\hat{O}$, $\hat{H}$, $\hat{N}$ all individuals play, respectively, strategies *O*, *H*, *N*.

Notice that figures 3a-3f illustrate that a bistable dynamic always holds in the edge of the simplex joining the stationary states $\hat{N}$ and $\hat{P}$, along which only strategies *N* and *P* are adopted. If the initial share of *N* players is high enough, then society will converge to the social poverty trap $\hat{N}$ where everyone chooses isolation. On the other hand, if the initial size of the population of *P* players is big enough, then society will reach the equilibrium $\hat{P}$ where people develop social relationships both by means of face-to-face interactions and the use of online networks, and SNS-mediated interactions are polite. However, if incivility spreads in online interactions, that is, if a positive share of individuals plays the *H* strategy (this occurs in the interior of the simplexes illustrated in figures 3a-3f), then the equilibrium $\hat{P}$ where all agents behave politely may cease to be attractive (this occurs in the regimes illustrated in figures 3e and 3f). Differently from the equilibrium $\hat{P}$, the equilibrium $\hat{O}$ where all agents play the *O* strategy –i.e. they do not rely on online networks to develop their social relations– is always locally attractive independently of the possible rise of online incivility (see figures 5a-5c).

*6.3. Dynamic regimes in the simplex S*

Let us remember that "almost all" the trajectories of system (3) approach an attractive stationary state belonging to one of the edges $S_N$, $S_O$, $S_P$, and $S_H$ of the simplex *S*. According to the analysis developed in the preceding two sections, no stationary state where three strategies coexist can be attractive under the system (3), in that no stationary state lying in the interior of the edges $S_N$, $S_O$, $S_P$, and $S_H$ is attractive for the trajectories in such edges. Only one stationary state where two strategies coexist can be attractive: the stationary state $(0, x_2^*, 1-x_2^*, 0)$, where all individuals play either the *H* or the *P* strategy. The other stationary states that can be attractive are the pure population states $\hat{O}, \hat{H}, \hat{P}$, and $\hat{N}$, where all individuals play only one strategy. Obviously, the stationary states $\hat{O}, \hat{H}, \hat{P}, \hat{N}$ and $(0, x_2^*, 1-x_2^*, 0)$ are attractive if and only if they are attractive in each of the edges to which they belong. The analyses in the preceding sections suggest that we distinguish between two cases: the case in which $\beta > -\delta$ and $\gamma < \varepsilon$ hold and that in which $\beta < -\delta$ and $\gamma > \varepsilon$ hold (see condition (2)).

*6.3.1. The case in which $\beta > -\delta$ and $\gamma < \varepsilon$ hold*

In this context, the stationary states $\hat{N}$, $\hat{O}$, and $\hat{P}$ are always attractive, while the stationary state $\hat{H}$ is attractive only if $\beta > \eta$. No other attractive stationary state can exist and, therefore, almost all the trajectories converge to one of the vertices of the simplex *S* (see Figure 1), where only one strategy is adopted. The payoff of each individual in the state $\hat{N}$ is always lower than in the other attractive stationary states. By the illustrative device adopted in Hirshleifer and Martinez Coll (1991), we can represent the edges $S_N$, $S_O$, $S_P$, and $S_H$ of *S* in the plane. The simplex *S* can be imagined as based on the triangle $\hat{O} - \hat{H} - \hat{P}$, while $\hat{N}$ is the upper vertex, that in which all strategies are extinct except for *N* (by drawing the edges in the 3-dimensional euclidean space, all the $\hat{N}$ vertices will come together). In Figure 6 we illustrate the dynamics in the edges of $S$ in a context in which all the vertices $\hat{O}$, $\hat{H}$, $\hat{P}$, and $\hat{N}$ are attractive.

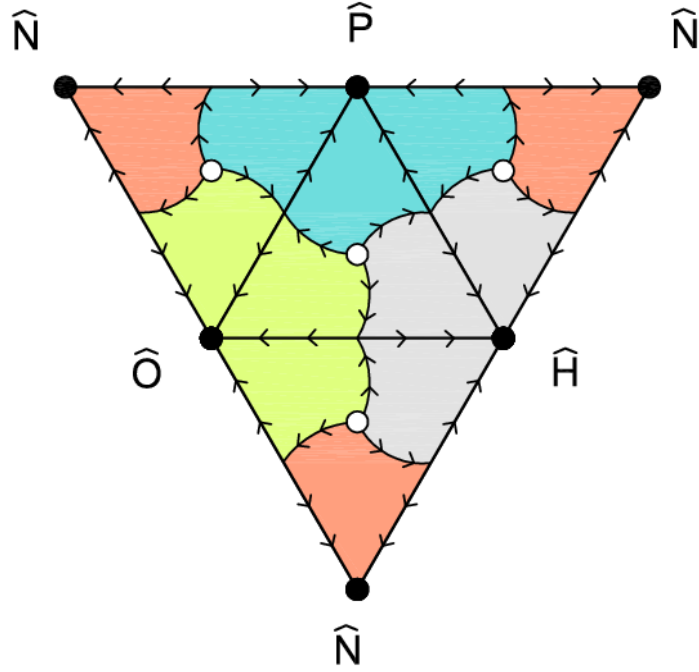

**Fig 6:** Dynamics in the edges of $S$ in a context in which all the vertices $\hat{O}$, $\hat{H}$, $\hat{P}$, $\hat{N}$ are attractive. The edges $S_N$, $S_O$, $S_P$, and $S_H$ of $S$ are represented in the plane. The simplex $S$ can be imagined as based on the triangle $\hat{O} - \hat{H} - \hat{P}$, while $\hat{N}$ is the upper vertex, that in which all strategies are extinct except for $N$ (by drawing the edges in the 3-dimensional euclidean space, all the $\hat{N}$ vertices will come together).

### 6.3.2. The case in which $\beta < -\delta$ and $\gamma > \varepsilon$ hold

In this context, the stationary states $\hat{O}$ and $\hat{N}$ are always attractive, while the stationary states $\hat{P}$ and $\hat{H}$ are never attractive. Furthermore, if condition (9) and the opposite of condition (7) hold, that is respectively:

$$\beta\varepsilon + \gamma\delta < 0$$

$$\eta < \frac{\beta\varepsilon + \gamma\delta}{\varepsilon - \gamma + \beta + \delta}$$

then there also exists the attractive stationary state $(0, x_2^*, 1-x_2^*, 0)$ where all individuals play either the $H$ or the $P$ strategy. In this context, we have:

$$\varepsilon = \Pi_P(0,1) > \Pi_H(x_2^*, 1-x_2^*) = \Pi_P(x_2^*, 1-x_2^*) > \Pi_H(1,0) = \beta$$

and therefore individuals' welfare in $(0, x_2^*, 1-x_2^*, 0)$ is lower than in the (non-attractive) stationary state $\hat{P}$ and higher that in the (non-attractive) stationary state $\hat{H}$; furthermore, the welfare in $(0, x_2^*, 1-x_2^*, 0)$ is higher than in $\hat{N}$.

In Figure 7 we illustrate the dynamics in the edges of $S$ in a context in which the vertices $\hat{O}$ and $\hat{N}$, and the coexistence stationary state $(0, x_2^*, 1-x_2^*, 0)$, are attractive.

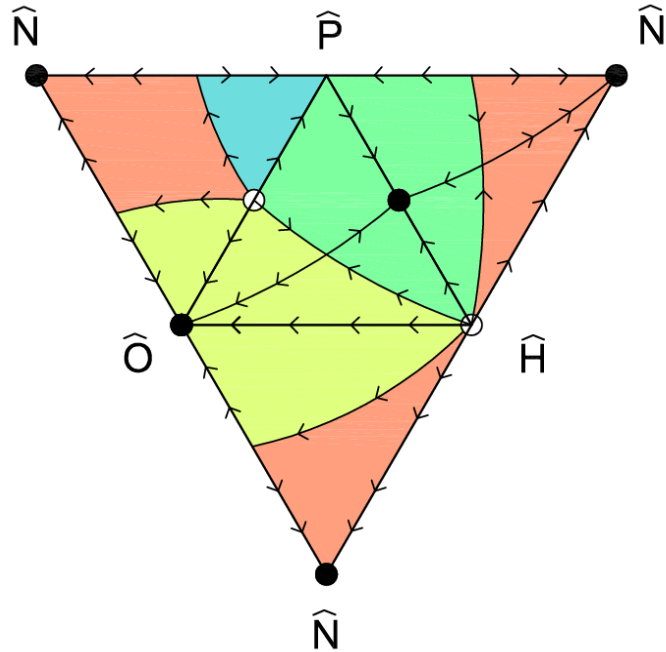

**Figure 7:** Dynamics in the edges of $S$ in a context in which the vertices $\hat{O}$ and $\hat{N}$, and the coexistence stationary state $(x_1, x_2, x_3, x_4) = (0, x_2^*, 1-x_2^*, 0)$ (where only strategies $H$ and $P$ are played), are attractive.

## 7. Discussion and conclusions

In this paper we have built an evolutionary game model to study the ecology of online and offline social interaction in a society where agents can develop their social relations by means of face-to-face encounters and interaction in online social networks. We have assumed that both offline and online social environments can become hostile due to forms of social decay, and that individuals can decide to isolate themselves as a self-protective strategy to cope with this.

The analysis of dynamics has shown that the spread of isolating self-protective behaviors could lead the economy to non-socially optimal stationary states that are Pareto dominated by others. For individuals, self-protective behaviors are rational in that they temporarily provide higher payoffs. However, their spread causes a generalized decrease in the payoffs associated with each strategy, which, in the long run, leads the economy to non-optimal outcomes.

Our model has shown that offline and online social decay worsens with the increase in the share of the population adopting the self-protective *N* strategy (entailing the choice of social isolation) and impolite *H* strategy (entailing the adoption of an uncivil behavior in online interactions).

The four pure population stationary states, where all individuals adopt the same strategy, can be simultaneously attractive. The destination of dynamics strictly depends on the initial distribution of the strategies in the population. This path dependence suggests that societies that are similar along a number of fundamental features can converge to different equilibria depending on their initial conditions. Given two countries with very similar trends in economic fundamentals, the social equilibria they converge to can be very different depending on some features of their initial social capital (e.g., attitudes towards peers and propensity to withdraw from social interaction) and the mechanisms of diffusion of SNS. This result calls for more careful measurement of the different facets of social capital as economic forces that might negatively evolve with economic growth, perhaps through technology adoption (as the mechanism shown in this paper).

The social poverty trap is always a locally attractive Nash equilibrium. The stationary states entailing positive levels of participation can be attractive depending on the configuration of parameters. When this happens, they always give higher payoffs than the social poverty trap.

The *H* and the *P* strategy can coexist when $\beta < -\delta$ and $\gamma > \varepsilon$. As discussed, this describes a situation in which the stakes involved in partner selection are higher for haters than for polite participants: when paired with the right partner (i.e., a *P* player), the hater is better off than another polite participant, whereas the former is worse off than the latter when paired with the wrong partner (i.e., another *H* player). Only under these conditions can the shared partner preference of *H* and *P* players (they both prefer to interact with another polite participant) lead to a mixed equilibrium with multiple strategies. It is worth noting that such shared preference can be present also in other scenarios (i.e., whenever $\gamma > \beta$ and $\varepsilon > -\delta$), but it is capable of supporting the coexistence of the *H* and *P* strategy at equilibrium only when the stakes of partner selection are higher for haters than for polite participants. This may have relevant policy implications: while affecting the negative stakes of partner selection for haters may be arduous (since it depends on their own behavior towards each other) and affecting their positive stakes is undesirable (making it more enjoyable for them to torment their victims would not be a commendable policy), it should be possible to intervene in the stakes of partner selection for polite participants. In particular, our results suggest that polite participants would be better off by caring *less* for partner selection, rather than more, since politeness can survive as a stable strategy in a world with a fair share of haters only if polite people care less about partner selection than haters do. Thus it would seem that Internet users engaged with haters need to heed the same advice Virgil gave Dante upon entering Hell: "Let us not speak of them, but do thou look and pass on".

Nonetheless, policies aimed at modifying individual payoffs might not be sufficient to prevent social poverty traps. From an institutional perspective, what could policy makers do to help people out of complete isolation and restore social interactions? Should governments intervene, or are there market forces that could be leveraged to do so? Antoci, Sacco and Vanin (2001) extensively argue for the need for complementary actions between governments and civil society. However, this model is pessimistic about the role of civil society; when a social trap forms, the whole population converges to the Pure Strategy equilibrium $\hat{N}$, without any convenient individual deviation. The dissemination of information on the existence of incivility online and the reasons why it can be a serious problem for society should be of primary concern for policy makers and SNS users alike. Therefore the government should probably enforce policies to prevent defensive self-isolating behaviors (e.g., school education on SNS and how to react to incivility) or to re-establish social connections (e.g., free public events, public goods with a social component). Future research should shed light on these issues.

# Mathematical appendix

It is easy to check that dynamics (6) can be written in the following form (see, e.g., Bomze 1983):

$$\dot{x}_i = x_i[e_i \cdot Ax - x \cdot Ax], \qquad i=1,2,3 \tag{12}$$

where $x$ is the vector $x = (x_1, x_2, x_3)$, $e_i$ is the vector of the canonical basis with $i$-entry equal to 1, and the others equal to 0, and $A$ is the payoff matrix:

$$A = \begin{pmatrix} \alpha & 0 & 0 \\ 0 & \beta & \gamma \\ 0 & -\delta & \varepsilon \end{pmatrix}$$

It is well-known that dynamics (12) does not change if an arbitrary constant is added to each column of $A$ (see, e.g., Hofbauer and Sigmund, 1988; p. 126). So we can replace matrix $A$, in equations (7), by the following *normalized* matrix $B$ with the first row made of zeros:

$$B = \begin{pmatrix} 0 & 0 & 0 \\ a & b & c \\ d & e & f \end{pmatrix} = \begin{pmatrix} 0 & 0 & 0 \\ -\alpha & \beta & \gamma \\ -\alpha & -\delta & \varepsilon \end{pmatrix}$$

We analyze the dynamics in the edge $S_N$ of the three-dimensional simplex $S$ where the $N$ strategy is not played by using Bomze's classification (1983) for two-dimensional replicator dynamic systems. The edge $S_N$ is a two-dimensional simplex that is invariant under replicator equations (3). We assume that parameters' values satisfy conditions (1) and (2), in particular:

$$\alpha, \delta, \varepsilon > 0$$
$$\beta > -\delta \text{ and } \gamma < \varepsilon, \text{ or } \beta < -\delta \text{ and } \gamma > \varepsilon \tag{13}$$

The parameters $\beta$ and $\gamma$ may be either positive or negative.

The vertices $\hat{O}$, $\hat{H}$, and $\hat{P}$ of the simplex $S_N$ correspond to the pure population states:

$$\hat{O} = (x_1, x_2, x_3, x_4) = (1,0,0,0)$$
$$\hat{H} = (x_1, x_2, x_3, x_4) = (0,1,0,0)$$
$$\hat{P} = (x_1, x_2, x_3, x_4) = (0,0,1,0)$$

where only strategies *O*, *H*, and *P* are respectively played. We shall indicate by $\hat{O}-\hat{H}$ the edge of $S_N$ joining $\hat{O}$ and $\hat{H}$ (where only strategies *O* and *H* are present in the population), by $\hat{O}-\hat{P}$ the edge joining $\hat{O}$ and $\hat{P}$ (where only strategies *O* and *P* are present), and by $\hat{H}-\hat{P}$ the edge joining $\hat{H}$ and $\hat{P}$ (where only strategies *H* and *P* are present). By *Int* $S_N$ we shall indicate the interior of the simplex $S_N$, that is the set in which all the strategies *O*, *H*, and *P* are played by strictly positive shares of the population.

In order to apply Bomze's classification, we make use of the same terminology introduced in Bomze (1983). By *an eigenvalue EV of a stationary state* we shall understand an eigenvalue of the linearization matrix around that stationary state. The term *EV* in direction of the vector *V* means that *V* is an eigenvector corresponding to that *EV*.

Let us observe first that the pure population states in which only one strategy is adopted by individuals, $\hat{O}$, $\hat{H}$ and $\hat{P}$, are always stationary states under replicator dynamics. Their stability properties are analyzed in the following proposition. For simplicity, the propositions in Bomze (1983) will be indicated as B# (so, e.g., B4 is Proposition 4 of Bomze's paper).

**Proposition 1** *The eigenvalue structure of the stationary states $\hat{O}$, $\hat{H}$, and $\hat{P}$ is the following:*

*(1) $\hat{O}$ has one eigenvalue with the sign of $a = -\alpha < 0$ in direction of $\hat{O}-\hat{H}$ and one eigenvalue with the sign of $d = -\alpha < 0$ in direction of $\hat{O}-\hat{P}$.*

*(2) $\hat{H}$ has one eigenvalue with the sign of $-b = -\beta$ in direction of $\hat{O}-\hat{H}$ and one eigenvalue with the sign of $e - b = -\beta - \delta$ in direction of $\hat{H}-\hat{P}$.*

*(3) $\hat{P}$ has one eigenvalue with the sign of $-f = -\varepsilon < 0$ in direction of $\hat{O}-\hat{P}$ and one eigenvalue with the sign of $c - f = \gamma - \varepsilon$ in direction of $\hat{H}-\hat{P}$.*

**Proof.** See B1.

Notice that: 1) The stationary state $\hat{O}$ is always (locally) attractive. 2) The stationary state $\hat{H}$ is attractive if $\beta > 0$, a saddle if $\beta < 0$ and $\beta > -\delta$, repulsive if $\beta < 0$ and $\beta < -\delta$. 3) The stationary state $\hat{P}$ is attractive if $\gamma < \varepsilon$, a saddle if $\gamma > \varepsilon$.

The following proposition concerns the stationary states on the interior of the edges of $S_N$.

**Proposition 2**

*(1) There is one stationary state in the interior of the edge $\hat{O}-\hat{H}$ if and only if (iff) $ab = -\alpha\beta < 0$ (i.e. $\beta > 0$), with eigenvalues having the sign of $-a = \alpha > 0$ in direction of $\hat{O}-\hat{H}$ and of:*

$$\frac{bd - ae}{b} = -\frac{\alpha}{\beta}(\beta + \delta)$$

*in direction of the interior of $S_N$. If $ab \geq 0$, then no stationary state exists in the interior of $\hat{O}-\hat{H}$.*

*(2) There always exists a unique stationary state in the interior of the edge $\hat{O}-\hat{P}$, with eigenvalues having the sign of $-d = \alpha > 0$ in direction of $\hat{O}-\hat{P}$ and of:*

$$\frac{af - cd}{f} = \frac{\alpha}{\varepsilon}(\gamma - \varepsilon)$$

*in direction of the interior of $S_N$.*

*(3) There always exists a unique stationary state in the interior of the edge $\hat{H}-\hat{P}$, with eigenvalues having the sign of:*

$$\frac{(e-b)(f-c)}{e-b+c-f} = \frac{(\beta+\delta)(\varepsilon-\gamma)}{(\beta+\delta)+(\varepsilon-\gamma)}$$

*in direction of $\hat{H}-\hat{P}$ and of:*

$$\frac{bf - ce}{e-b+c-f} = -\frac{\beta\varepsilon + \gamma\delta}{(\beta+\delta)+(\varepsilon-\gamma)}$$

*in direction of the interior of $S_N$.*

**Proof.** Apply B2 and B5 taking into account that, according to assumption (2), $\beta + \delta$ and $\varepsilon - \gamma$ have the same sign, and $\beta \neq -\delta$ and $\varepsilon \neq \gamma$.

The remaining proposition concerns the stationary states in the interior of $S_N$, where all the strategies *O*, *H*, and *P* coexist.

**Proposition 3** *There is a unique stationary state in $\text{Int } S_N$ if the expressions:*

$$bf - ce = \beta\varepsilon + \gamma\delta \qquad ae - bd = \alpha(\beta + \delta) \qquad cd - af = \alpha(\varepsilon - \gamma) \tag{14}$$

*are all either strictly positive or strictly negative. In the remaining cases, there are not stationary states in Int $S_N$.*

**Proof.** Apply B6.

Notice that, according to condition (2), $\beta + \delta$ and $\varepsilon - \gamma$ have the same sign; furthermore, $\beta \neq -\delta$ and $\varepsilon \neq \gamma$ always hold. Consequently, an interior stationary state exists if $\beta\varepsilon + \gamma\delta$ has the same sign of $\beta + \delta$ and $\varepsilon - \gamma$.

According to Propositions 1-3, the dynamic regimes that may be observed in the edge $S_N$ are the following:

**1)** Case $\beta > 0$ (and therefore $\beta > -\delta$) and $\gamma < \varepsilon$. In this case, all the vertices $\hat{O}$, $\hat{H}$, and $\hat{P}$ are attractive. There exist stationary states in the interior of the edges $\hat{O}-\hat{H}$ and $\hat{O}-\hat{P}$, and they are saddles with unstable manifolds belonging to the edges; there exists a stationary state in the interior of $\hat{H}-\hat{P}$, which is a saddle (with unstable manifold belonging to the edge) if $\beta\varepsilon + \gamma\delta > 0$ and repulsive if $\beta\varepsilon + \gamma\delta < 0$. Finally, the stationary state in *Int* $S_N$ exists if $\beta\varepsilon + \gamma\delta > 0$. Figures 2a and 2b illustrate, respectively, the case $\beta\varepsilon + \gamma\delta > 0$ and the case $\beta\varepsilon + \gamma\delta < 0$ (they correspond, respectively, to phase portraits number 7 and 35 of Bomze's classification).

**2)** Case $\beta < 0$ and $\gamma < \varepsilon$ (and therefore $\beta > -\delta$, by conditions (2)). In this case, the vertices $\hat{O}$ and $\hat{P}$ are locally attractive, while $\hat{H}$ is a saddle point with stable manifold belonging to the edge $\hat{H}-\hat{P}$. No stationary state exists in the interior of the edge $\hat{O}-\hat{H}$; there exists a saddle point in the interior of the edge $\hat{O}-\hat{P}$ (with unstable manifold belonging to the edge); there exists a stationary state in the interior of $\hat{H}-\hat{P}$, which is a saddle (with unstable manifold belonging to the edge) if $\beta\varepsilon + \gamma\delta > 0$, while it is repulsive if $\beta\varepsilon + \gamma\delta < 0$. Finally, the stationary state in *Int* $S_N$ exists if $\beta\varepsilon + \gamma\delta > 0$. Figures 2c and 2d illustrate, respectively, the case $\beta\varepsilon + \gamma\delta > 0$ and the case $\beta\varepsilon + \gamma\delta < 0$ (they correspond, respectively, to phase portraits number 9 and 37 of Bomze's classification).

3) Case $\beta < 0$ and $\gamma > \varepsilon$ (and therefore $\beta < -\delta$, by conditions (2)).[14] In this case, the vertex $\hat{O}$ is attractive, $\hat{H}$ is repulsive and $\hat{P}$ is a saddle with unstable manifold belonging to the edge $\hat{H}-\hat{P}$. No stationary state exists in the interior of the edge $\hat{O}-\hat{H}$; a repulsive stationary state exists in the interior of the edge $\hat{O}-\hat{P}$; there exists a stationary state in the interior of $\hat{H}-\hat{P}$, which is a saddle (with stable manifold belonging to the edge) if $\beta\varepsilon+\gamma\delta>0$, while it is attractive if $\beta\varepsilon+\gamma\delta<0$. Finally, the stationary state in $Int\ S_N$ exists if $\beta\varepsilon+\gamma\delta<0$ and it is a saddle point. Figures 2e and 2f illustrate, respectively, the case $\beta\varepsilon+\gamma\delta<0$ and the case $\beta\varepsilon+\gamma\delta>0$ (they correspond, respectively, to phase portraits number 11 and 36 of Bomze's classification).

---
[14] The case $\beta > 0$ and $\gamma > \varepsilon$ is excluded by conditions (1).